\documentclass[floatfix,aps,prl,showpacs,amsmath,nofootinbib,preprintnumbers,twocolumn]{revtex4}

\def\b{\begin{equation}}
\def\e{\end{equation}}

\def\ap{\alpha '}
\def\d{\Lambda}

\def\[{\left [}
\def\]{\right ]}
\def\({\left (}
\def\){\right )}

\begin{document}

\title{Weak gravity conjecture in the asymptotical dS
and AdS background}

\author{Qing-Guo Huang}\email{huangqg@itp.ac.cn}
\affiliation{Interdisciplinary Center of Theoretical Studies,
Academia Sinica, Beijing \ 100080, China}

\author{Miao Li}\email{mli@itp.ac.cn}
\affiliation{Institute of Theoretical Physics, Academia Sinica,
Beijing \ 100080, China \\
Interdisciplinary Center of Theoretical Studies, Academia Sinica,
Beijing \ 100080, China}

\author{Wei Song}\email{wsong@itp.ac.cn}
\affiliation{Institute of Theoretical Physics, Academia Sinica,
Beijing \ 100080, China}

\date{\today}

\begin{abstract}

The cosmological observations provide a strong evidence that there
is a positive cosmological constant in our universe and thus the
spacetime is asymptotical de Sitter space. The conjecture of gravity
as the weakest force in the asymptotical dS space leads to a lower
bound on the U(1) gauge coupling $g$, or equivalently, the positive
cosmological constant gets an upper bound $\rho_V \leq g^2 M_p^4$ in
order that the U(1) gauge theory can survive in four dimensions.
This result has a simple explanation in string theory, i.e. the
string scale $\sqrt{\alpha '}$ should not be greater than the size
of the cosmic horizon. Our proposal in string theory can be
generalized to U(N) gauge theory and gives a guideline to the
microscopic explanation of the de Sitter entropy. The similar
results are also obtained in the asymptotical anti-de Sitter space.

\end{abstract}

\maketitle


Particle physicists have told us the low energy physics is perfectly
described by the standard model. Recent cosmological observations
provide a strong evidence that there is a positive cosmological
constant in our universe. As the only well formulated candidate for
quantum gravity, string theory shall be connected with the phenomena
in our universe. A central topic along the line in string theory is
to answer why the theory for the low energy phenomena is the
standard model, why there is a cosmological constant, and why it is
the value we observed.

String theory is only consistent in ten dimensions in order to
cancel the conformal anomaly on the string world sheet. To
understand four dimensional physics in our real world, we must
compactify string theory on some manifolds. Recent developments for
the flux compactifications \cite{fc}, however, suggest that a huge
number of at least semi-classically consistent string vacua emerge
in string theory, named string landscape \cite{ls}.  It may or may
not provide an opportunity for us to explore the specific low energy
phenomena in the experiments from the viewpoint of string theory.
Given the numerous ``vacua" in the string landscape, the most urgent
problem is to find a reliable vacuum selecting principle.

 In \cite{cv}, Vafa proposed that
self-consistence of a quantum theory of gravity offers a way to pick
out which effective field theories can arise. Many vacua in the
string landscape, although consistent semi-classically, are actually
inconsistent on the quantum level, called swampland in \cite{cv}.
Self-consistent landscape is surrounded by the swampland. Then the
problem is translated into finding the criteria to pick out really
self-consistent landscape from the swampland. More criteria for
self-consistent effective field theory have been proposed in
\cite{aadnr,amnv}. Eventually we expect that more and more
consistent conditions will be found and the range for otherwise free
parameters will be narrowed down and the predictions of string
theory can be checked in the experiments.

The authors in \cite{aadnr, amnv} proposed some criteria for the
self-consistent effective field theory in four-dimensional
asymptotical flat spacetime. However, there are a huge number of
string vacua with positive or negative cosmological constants in
string landscape. What is more, our universe is asymptotical de
sitter space which is favored by recent cosmological observations.
So we are motivated to explore the criterion for self-consistent
effective field theory in asymptotical de Sitter and Anti-de Sitter
background. On the other hand, the well-defined string theory and
quantum field theory in asymptotical dS space are still unknown. But
we try to explore them heuristically in this note.

In \cite{amnv}, gravity is conjectured as the weakest force. It is
the claim that for a $U(1)$ theory there exists a charged particle
whose mass is smaller than its charge in some appropriate unit. The
conjecture is supported by string theory and some evidence involving
black holes and symmetries. The conjecture leads to an intrinsic
bound on the UV cutoff $\d$ for a consistent U(1) gauge field theory
in asymptotical flat four dimensional spacetime which takes the form
\b{\label{fbg} \d \leq g M_p\sim g/\sqrt{G},}\e where $g$ is the
U(1) gauge coupling. An intrinsic radius of curvature appears in the
asymptotical de Sitter and Anti-de Sitter background. An infrared
cut-off of the effective field theory should be greater than, if not
equal to, the radius of the background. Moreover, the UV cut-off of
the effective field theory should be smaller than the curvature
radius, otherwise the full quantum gravity or string theory must be
invoked to describe the situation. We shall show that the latter
condition combined with the conjecture of gravity as the weakest
force yields a lower bound on the U(1) gauge theory; or
equivalently, an upper bound on the cosmological constant. We also
propose a heuristic explanation in string theory for this bound of
the coupling constant.

Let us start with the metric of Schwarzschild-de Sitter solution in
four dimensional spacetime \b{\label{msds} ds^2=-f(r)dt^2+
f^{-1}(r)dr^2+r^2d\Omega_2^2, }\e with \b{ \label{fr}
f(r)=1-\frac{2Gm}{r}-\frac{r^2}{L^2}, }\e where $G$ is the Newton
constant and $L=\sqrt{3 / (8\pi G \rho_V)}$ is the size of the pure
de Sitter space with a positive cosmological constant $\rho_V$. For
a U(1) gauge theory, the mass scale of the minimally charged
monopole is $\d/g^2$ and its size is of order $1/\d$, where $\d$ is
the cutoff of the field theory. The size of the black hole horizon
$r_-$ and cosmic horizon $r_+$ satisfies \b{r_\pm^3-L^2r_\pm+2G{\d
\over g^2}L^2=0.}\e Requiring that this monopole is not black and
smaller than the cosmic horizon, namely $r_-\leq {1 \over \d}\leq
r_+$, yields \b{{1 \over \d^3}-{L^2 \over \d}+2G{\d \over
g^2}L^2\leq 0}\e or, \b{\label{pdg} \d^4-{g^2 \over 2G}\d^2+{g^2
\over 2GL^2}\leq 0.}\e Demanding that there is a solution for the
inequality (\ref{pdg}) yields \b{\label{bg} g\geq {\sqrt{8G} \over
L},}\e or equivalently, \b{\label{bcc} \rho_V \leq g^2/G^2 \sim
g^2M_p^4. }\e If there is a very weak U(1) gauge theory with gauge
coupling $g\sim 10^{-60}$, the cosmological constant is roughly the
same as that we observed. Solving the inequality (\ref{pdg}), we
find a bound on the cutoff for U(1) gauge theory which takes the
form \b{\label{pdc} {g \over 2\sqrt{G}}\sqrt{1-\sqrt{1-{8G \over
g^2L^2}} } \leq \d \leq {g \over 2\sqrt{G}}\sqrt{1+\sqrt{1-{8G \over
g^2L^2}} }. }\e For a fixed gauge coupling, in the limit with
$\rho_V \rightarrow 0$ or $L \rightarrow \infty$, eq. (\ref{pdc}) is
just the same as eq. (\ref{fbg}). When gauge coupling goes to its
lower bound, the UV cutoff for this U(1) gauge field theory is $\d
\sim g/\sqrt{G}\sim 1/L$.

Surprisingly, eq.(\ref{bg}) shows that a positive cosmological
constant induces a lower bound on the U(1) gauge coupling. Or
equivalently, the positive cosmological constant can not be
arbitrarily large in order that a consistent U(1) gauge theory can
survive. The most important input at this point is the requirement
that the size of the minimally charged monopole is not larger than
the cosmic horizon in the asymptotical de Sitter space. This is also
the condition for us to trust the above estimates about the mass
scale and the size of the monopole.

There is a simple physical explanation of eq.(\ref{bg}). $1/\d$ is
roughly the shortest physical length for the U(1) gauge field
theory. It is natural to demand $1/\d$ be no larger than the size of
cosmic horizon, namely $\d>1/L$. Together with $\d \le gM_p$, this
directly leads to eq.(\ref{bg}) and (\ref{bcc}).

Another heuristic consideration leading to eq.(\ref{bg}) is the
following. In order that there is no naked singularity in the
space-time, the mass of the minimally charged monopole is not
greater than the mass parameter of the Nariai Black hole $L/G$,
namely \b{\label{bn} {\d \over g^2}\leq {L\over G}, \quad \hbox{or}
\quad \d \leq {g^2L \over G}. }\e On the other hand, the size of the
monopole $1/\d$ should not be larger than the size scale of the
cosmic horizon $L$; or equivalently, $\d \geq 1/L$. Substituting
this relationship into eq. (\ref{bn}), we obtain eq. (\ref{bg}) and
(\ref{bcc}) again.

In the more formal derivation using the metric (2), we did not introduce
in $f(r)$ the contribution of the magnetic charge which is roughly
$Gg^{-2}r^{-2}$, this term is smaller than $GMr^{-1}$ if the horizon
size is larger than $1/\d$. Or it is larger than $GMr^{-1}$ if the
horizon size is smaller than $1/\d$, in this case we obtain the condition
$\d\le gM_p$. Thus, if we include the term $Gg^{-2}r^{-2}$ in the above
formal discussion, we will end up with inequalities similar to (9).

In \cite{amnv}, the authors argued that the absence of global
symmetries in quantum gravity requires that the field theory
description should break down in the limit  $g\rightarrow 0$, since
the symmetry can be identified as a global symmetry. The way to
avoid this problem in \cite{amnv} is the UV cutoff also goes to zero
when $g\rightarrow 0$. In a asymptotical de Sitter space, an
intrinsic lower bound on the gauge coupling is induced by the size
of the cosmic horizon (\ref{bg}), which shows that the effective
gauge field theory already breaks down before taking the limit
$g\rightarrow 0$. The gauge coupling characterizes the strength of
the local interaction. Eq. (\ref{bg}) implies that the size of the
system can affect the local interaction in quantum theory.

We pause to discuss the most important premise in our discussion, namely
a minimally charged monopole should not be a black hole. The following
reasoning not only applies to an asymptotic Minkowski space, it also applies
to an asymptotic de Sitter space. Imagining that a minimal charged
monopole is indeed a black hole, thus it will Hawking radiate (the horizon size
is greater than the field theory UV cut-off which in turn is greater
than the Planck scale). During the
radiation process, neutral particles as well as charged particles can be
radiated. If only neutral particles are radiated, the black hole's mass
becomes smaller while its magnetic charges remains the same, this implies
that there exists monopoles with smaller mass.
If charges are radiated, these charges must be smaller than that of the original monopole,
this contradicts our assumption. Of course the above argument does not
apply to a general black hole with magnetic charge, since it may be formed
of many minimally charged monopoles and other matter, thus is neither minimally
charged nor with smaller mass. \cite{amnv} derives the weak gravity inequality
using the absence remnants, which is valid also in an asymptotic de Sitter
space.

We now switch to string theory. Consider the brane world scenario in
Type IIB string theory. The tension of the D3-brane $T_3\sim
M_s^4/g_s$ is taken as the effective cosmological constant on the
brane and the U(1) gauge coupling is related to the string coupling
$g_s$ by $g\sim g_s^{1/2}$. According to eq. (\ref{bcc}), we obtain
a constraint on the string scale and the string coupling, namely
\b{\label{sgp} M_s^2 \leq g_s M_p^2. }\e Note that the string theory
in four dimensions is reduced from ten dimensions. For toroidal
compactification, if the average size of the extra dimension is $R$,
the Planck scale in four dimensions is \b{\label{msg} M_p^2\sim
R^6M_s^8/g_s^2=(RM_s)^6M_s^2/g_s^2. }\e Requirement (\ref{sgp})
implies \b{\label{grms} g_s\leq (RM_s)^6. }\e In general we assume
$RM_s>1$; otherwise, we switch to a T-dual description. For weakly
coupled string theory $g_s\leq 1$, this condition is always
satisfied. The constraint on the string coupling is quite loose.

In string theory, we can find a simple explanation about the lower
bound on the gauge coupling or string coupling in the asymptotical
de Sitter space. Only when the length of string $\sqrt{\ap}$ is
shorter than the size of the cosmic horizon, the stringy effects can
be ignored and the description of the effective field theory is
reliable. This is just the condition that the Hawking temperature is
lower than the string Hagedorn temperature. Thus we require
\b{\label{stl} \sqrt{\ap}\leq L, \quad \hbox{or} \quad \rho_V \leq
{1 \over G\ap}. }\e In four dimensions Newton's constant is related
to the string coupling and string length square $\ap\sim 1/M_s^2$ by
\b{\label{ggp} G\sim g_s^2\ap}\e up to a coefficient which depends
on the compactification.  Substituting eq. (\ref{ggp}) into eq.
(\ref{stl}), we obtain \b{\label{bccs} \rho_V \leq g_s^2 M_p^4. }\e
In an asymptotical de Sitter space string coupling can not be
arbitrarily weak. For a given string coupling, an upper bound on the
cosmological constant appears; Above the bound, the effective gauge
field theory on the brane breaks down.

Even though a well-defined string theory in asymptotical de Sitter
is still unknown, the above discussions provide a useful constraint
on possible realizations of de Sitter space. We now  re-investigate
the brane world scenario in Type II B string theory more carefully.
Assume the string theory in four dimensions be reduced from ten
dimensions by toroidal compactification. Eq. (\ref{ggp}) is modified
to  \b{\label{cggp} G\sim g_s^2 \ap (RM_s)^{-6}}\e Thus eq.
(\ref{bccs}) takes the form \b{\label{dbccs} \rho_V \leq g_s^2 M_p^4
(RM_s)^{-6}. }\e Identifying $\rho_V$ with $T_3\sim M_s^4/g_s$ and
using eq. (\ref{cggp}) we find $g_s\leq (RM_s)^6$ which is exactly
the same as eq. (\ref{grms}). This result obtained in string theory
exactly matches the result obtained in the effective field theory.

We can go one step further in string theory. The field theoretical
argument can not be generalized to the Non-Abelian gauge field
theory, while the string theory argument can. Consider a stack of
$N$ D3-branes. The fields of the open string theory are in the
adjoint representation of SU($N$). For a stack of D3-brane, the
effective cosmological constant becomes $NT_3\sim NM_s^4/g_s$. In
this case we simply  obtain the constraint on the string coupling as
\b{\label{ngs} g_sN\leq (RM_s)^6. }\e The combination of the string
coupling and $N$ is nothing but t' Hooft coupling. This is to be
expected. For fixed string coupling and size of the extra
dimensions, an upper bound on the rank of the gauge group is
obtained. Eq.(19) can be obtained from eq.(8) provided $g^2$ in
eq.(8) is replaced by $g^2N$.

The above discussions in strng theory can be generalized to diverse
dimensions. For simplicity, we investigate a stack of $N$ D9-brane.
The Hubble parameter $H$ on the brane takes the form \b{\label{hpb}
H\sim \sqrt{G_{10}T_9 N}\sim {1 \over l_s}\sqrt{g_sN}, }\e where the
ten-dimensional Newton constant is given by $G_{10}\sim g_s^2l_s^8$
and $T_9\sim 1/(g_s l_s^{10})$. Requiring $\sqrt{\ap}\leq H^{-1}$
yields \b{\label{tgs}g_sN\leq 1.}\e The 't Hooft coupling must be
not greater than 1 in order that the gauge field theory on the brane
is effective. The entropy of de Sitter space on the brane is
\b{S\sim {1 \over H^8 G_{10}}\sim {1 \over (g_sN)^6}N^2.}\e For
$g_sN\leq 1$, $S\geq N^2$. This is a reasonable result since the
number of adjoint fields is no less than $N^2$. Recall the argument
about de Sitter entropy in \cite{fs}. The authors considered a
system of $N$ unstable D9-brane in Type II A string theory. The
basic requirement for eternal inflation is that the Hubble time
$H^{-1}$ should be larger than the time scale for the open string
tachyon to fall off the top of its potential $l_s$, which yields
$g_s N \geq 1$ by using eq. (\ref{hpb}). The gauge field theory on
the brane breaks down for eternal inflation unless $g_sN\sim 1$. For
fixed t' Hooft coupling $g_sN\sim 1$, we can take $N\rightarrow
\infty$ and $g_s\rightarrow 0$ without any need for closed string
quantum corrections.  Now the de Sitter entropy is just the square
of the number of branes $N^2$. This provides a tentative calculation
for the de Sitter entropy in string theory. We hope we can work out
the details on this argument in the future. This is different with
the case AdS/CFT where we require t' Hooft coupling is much greater
than one in order that we can trust the geometry \cite{agmoo}.

In an asymptotical Anti-de Sitter space-time, the similar results
can be obtained. The metric of Schwarzschild Anti-de Sitter solution
in four dimensional spacetime takes the form
\b{\label{msds}ds^2=-f(r)dt^2+ f^{-1}(r)dr^2+r^2d\Omega_2^2, }\e
with \b{ \label{fr}f(r)=1-\frac{2Gm}{r}+\frac{r^2}{L^2}, }\e where
$L=\sqrt{-3 / (8\pi G \rho_V)}$ is the size of the anti-de Sitter
space with a negative cosmological constant $\rho_V$. The radius of
the black hole $r_{bh}$ satisfies \b{r_{bh}^3+L^2r_{bh}-2GmL^2=0.
}\e Requiring that the minimally charged monopole should not be
black yields \b{{1\over \d^3}+{L^2 \over \d}-2G{\d \over g^2}L^2\geq
0, }\e or equivalently, \b{\label{apdg} \d^4-{g^2 \over
2G}\d^2-{g^2\over 2GL^2}\leq 0. }\e Solving this inequality, we
obtain the bound on the intrinsic UV cutoff for the U(1) gauge field
theory, namely \b{\label{apdc} \d \leq {g \over
2\sqrt{G}}\(1+\sqrt{1+{8G \over g^2L^2}} \)^{1/2}. }\e On the other
hand, we also require that the minimal physical length $1/\d$ should
be shorter than the radius of anti-de Sitter background; otherwise,
the gauge field theory breaks down. Thus $g\geq \sqrt{G}/L$ or
$|\rho_V| \leq g^2M_p^4$. With the viewpoint of string theory, the
similar results are also obtained.

In \cite{lsw}, the authors generalized the arguments in four
dimensions in \cite{amnv} to lower dimensions. Our previous
discussions can also be used to investigate the cases in lower
dimensions and the similar results are obtained.

To summarize, we have investigated the constraints on the effective
gauge field theory in an asymptotical de Sitter and an anti-de
Sitter background. But string theory still survives when the
constraints are violated. A lower bound on the gauge coupling
results from the requirement that the shortest length for the
effective gauge field theory should be shorter than the radius of
the background curvature. This result has a simple explanation in
string theory. The discusisons in string theory can be generalized
to diverse dimensions and the non-Abelian gauge field theory.

We also want to stress that we don't provide any concrete example to
show how certain theories in certain de Sitter space cannot arise in
string theory. We only say that there is no local field theory
description for length scales shorter than the de Sitter radius if
the latter itself is shorter than the string scale.

Acknowledgement. The work of QGH is supported by a grant from NSFC,
a grant from China Postdoctoral Science Foundation and and a grant
from K. C. Wang Postdoctoral Foundation. The work of ML and WS is
supported by a grant from CAS and a grant from NSFC.




\begin{thebibliography}{99}
\frenchspacing

\bibitem{fc}
S.B. Giddings, S. Kachru and J. Polchinski, Phys.Rev.D
66(2002)106006, hep-th/0105097; S. Kachru, R. Kallosh, A. Linde and
S.P. Trivedi, Phys.Rev.D 68(2003)046005.

\bibitem{ls}
L. Susskind, hep-th/0302219.

\bibitem{cv}
C. Vafa, hep-th/0509212.

\bibitem{amnv}
N. Arkani-Hamed, L. Motl, A. Nicolis and C. Vafa, hep-th/0601001.

\bibitem{aadnr}
A. Adams, N. Arkani-Hamed, S. Dubovsky, A. Nicolis, R. Rattazzi,
hep-th/0602178.

\bibitem{fs}
B. Freivogel and L. Susskind, Phys.Rev.D 70(2004)126007,
hep-th/0408133.

\bibitem{agmoo}
O. Aharony, S. S. Gubser, J. Maldacena, H. Ooguri and Y. Oz,
hep-th/9905111.

\bibitem{lsw}
M. Li, W. Song and T. Wang, hep-th/0601137.


\end{thebibliography}
\end{document}